\documentclass[a4paper,11pt]{article}
\usepackage{pos}
\usepackage{amsmath}
\usepackage{gensymb}
\usepackage{caption}
\usepackage{subcaption}
\usepackage{lineno}

\title{Spectral and Energy Morphology Analysis Study of HAWC J2031+415}
 \ShortTitle{Spectral and Energy Morphology Analysis Study of HAWC J2031+415}

\author*[a]{Ian Herzog}

\affiliation[a]{Michigan Technological University,\\
  1400 Townsend Drive, Houghton, USA}

\forColl{HAWC} 

\emailAdd{igherzog@mtu.edu}

\abstract{The Cygnus Cocoon region is a complex region containing an OB star cluster that is prominent in the TeV energy range. Located in this region is 3HWC J2031+415, a significant TeV gamma-ray source whose emission is possibly associated with 2 components, the Cygnus OB2 star cluster and a pulsar wind nebula (PWN). In this work, several modelling methods are presented to best describe the emission. These models disentangle emission believed to be from the Cocoon and isolate the component emitted by the probable PWN. I will present several spectral models to describe the emission of the probable PWN using the latest data set from the High-Altitude Water Cherenkov (HAWC) observatory.  Furthermore, I will present an energy morphology study of the PWN component of 3HWC J2031+415 in distinct energy bands.}

\FullConference{37$^{\rm{th}}$ International Cosmic Ray Conference (ICRC 2021)\\
		July 12th -- 23rd, 2021\\
		Online -- Berlin, Germany}

\begin{document}
\maketitle

\section{Introduction}

The first detection of a TeV source close to 3HWC J2031+415 was by HEGRA in 2002 and it was unique in that it had no obvious counterpart in lower energy ranges [7].  Designated TeV J2032+4130, they performed a point source fit to its emission and placed it firmly in the Cygnus Cocoon region. 

The first detection of a TeV source close to 3HWC J2031+415 was TeV J2032+4130, an extended very high energy source detected by HEGRA in 2002 [7].  Located in the Cygnus Cocoon region (henceforth referred to as the Cocoon) in the constellation Cygnus, it was unique in that there was no obvious counterpart in lower energy ranges.  A later study by VERITAS resulted in an extended asymmetric Gaussian fit being determined with data between ~.5 to ~30 TeV [6].  Furthermore, they hypothesized that the PWN is powered by PRS J2032+4127, a binary pulsar located near the source of the emission [2,6]. As a result of this hypothesis, they predicted a cutoff in the spectrum at a few tens of TeV [6].

The latest work on TeV J2032+4130 comes from the HAWC observatory in both the 3rd HAWC catalog, which gave it the name 3HWC J2031+415, and a dedicated analysis on the Cocoon region, calling it HAWC J2031+415 [1, 4].  The publication on the Cocoon resolved 3HWC J2031+415 into two distinct sources: the TeV continuation of the Cygnus Cocoon (HAWC J2030+409) and HAWC J2031+415 the PWN previously detected by VERITAS [4].  That study focused on isolating the Cocoon emission and as such they modelled the PWN and subtracted it out of the region.  This work will explore HAWC J2031+415's emission and resolve its energy morphology.

\section{Instrument and analysis}

The HAWC observatory is a wide field of view TeV gamma-ray observatory located at Sierra Negra in Mexico at an elevation of 4100 meters [3]. The main array consists of 300 water Cherenkov detectors that covers approximately 22,000 m$^2$ and is sensitive to gamma-rays in the range of 300 GeV - 100 TeV and beyond [1,3].  A secondary array called the outriggers is currently being implemented and contains an additional 345 smaller tanks and increasing the effective area to 100,000 m$^2$.  Each main array tank contains 4 photomultiplier tubes (PMT) that capture Cherenkov light emitted by high energy particles entering the tanks.  These events are collected and background and data maps are constructed from them [1,3].

HAWC data is binned using 3 methods: fhit, ground parameter, and the neural network [1,3].  The fhit binning method (bins 1-9) takes the percent of PMT's triggered by a shower and is efficient at low energies; at above 10 TeV much of the sensitivity is lost as the whole array is triggered.  A further refinement that was developed was the "ground parameter" energy estimator. This method divides each fhit bin into 12 quarter decade bins (a-l) and results in much higher sensitivity above 10 TeV.  The reconstruction method takes the charge density 40 meters from the shower axis to determine various binning parameters [3].  The neural network method trains a network to reconstruct the shower using a variety of parameters and is not considered for this analysis.  This study will use 1343 days of ground parameter data with reconstructed energies > 1 TeV. 

\section{Modelling the region}

Shown in Figure 1a is the initial significance map of the region of interest (ROI) considered for this analysis.  3HWC J2031+415 is the second brightest source in the Cocoon region and is detected with a significance of 23.6 $\sigma$ at the location RA = 307.93$\degree$, Dec = 41.51$\degree$ [1].  As previously mentioned, 3HWC J2031+415 can be resolved as HAWC J2030+409 and HAWC J2031+415.  Furthermore an additional source, 3HWC J2020+403, also known as the Gamma Cygni supernova remnant, will also be considered for this analysis [12].  The locations, morphologies, and the nearest associated sources are given in Table 1.  The ROI is a 6 degree circle centered at RA=306.5$\degree$,  Dec=40.7$\degree$, with a 2 degree mask placed on the brightest source in the region, 3HWC J2019+367, and is shown in Figure 1a.

\begin{figure}[t]
\centering
\begin{subfigure}{.45\textwidth}
  \centering
  \includegraphics[width=1\linewidth]{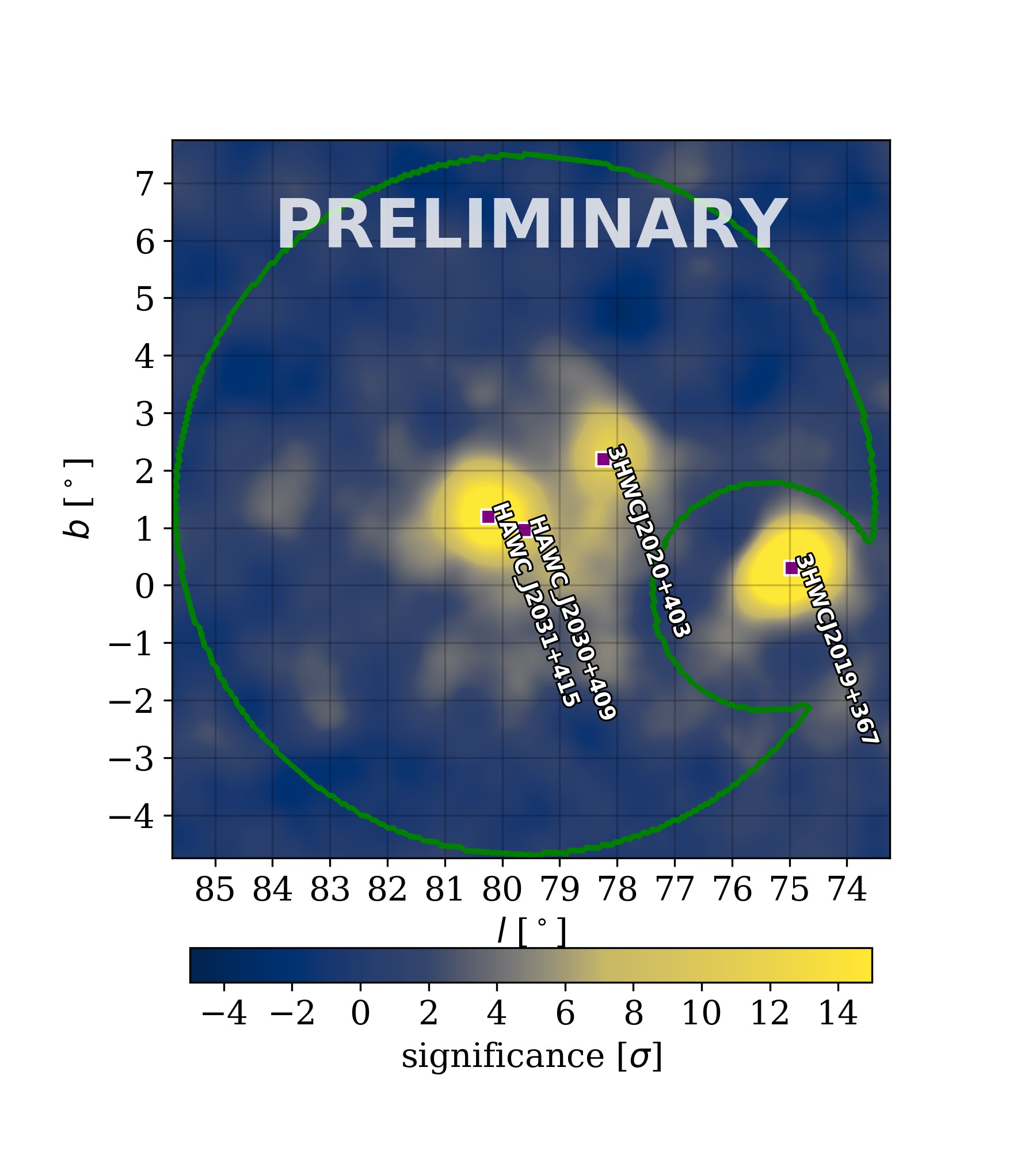}
  \caption{Significance map of the Cocoon region.  This shows the excess before removing 3HWC J2020+403 and HAWC J2030+409.}
  \label{fig:sub1}
\end{subfigure}\hfill
\begin{subfigure}{.45\textwidth}
  \centering
  \includegraphics[width=1\textwidth]{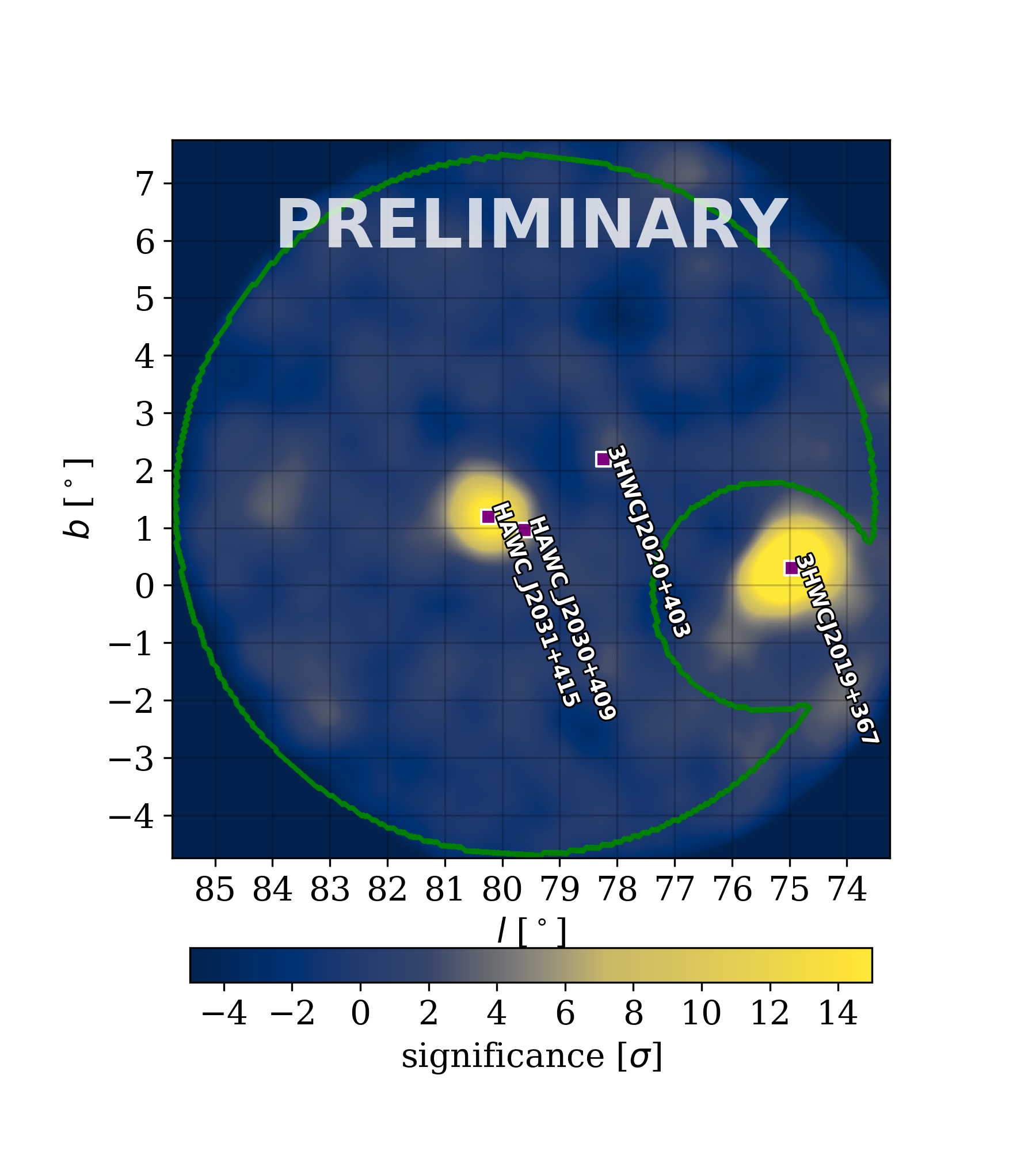}
  \caption{Significance map after subtracting 3HWC J2020+403 and HAWC J2030+409 as discussed in Section 3.1.  There is a significant excess after subtraction.}
  \label{fig:sub2}
\end{subfigure}
\caption{Significance maps of the Cocoon region.  Both maps are made with a 0.5$\degree$ extension and index of -2.7 assumptions and use 1343 days of ground parameter HAWC data with reconstructed energy > 1 TeV.  The green contour shows the ROI considered.}
\label{fig:test}
\end{figure}

\begin{table}[t]
\centering
 \begin{tabular}{||c c c c c||} 
 \hline
 Source & R.A [deg] & Dec. [deg] & Morphology & Source Association \\ [0.5ex] 
 \hline\hline
 3HWC J2020+403 & 305.27 & 40.50 & 0.63 (fixed) $\degree$ Gaussian width & Gamma Cygni SNR \\
 HAWC J2030+409 & 307.65 & 40.93 & 2.18$\pm$ 0.17 $\degree$ Gaussian width & Cygnus Cocoon \\
 HAWC J2031+415 & 307.82 & 41.51 & 0.27$\pm$ 0.02 $\degree$ Gaussian Width & VER J2031+415 PWN \\[1ex] 
 \hline
 \end{tabular}
\caption{The sources considered with location, morphology, and their associated counterparts [4, 6]}
\end{table}

\subsection{Fitting the region}

A multi-source likelihood fitting method is incorporated using the HAL \footnote{https://github.com/threeML/hawc\_hal} and 3ML framework [11].  All 3 sources are modelled as extended Gaussians and are initially fitted simultaneously.  Both 3HWC J2020+403 and HAWC J2030+409 are simple power laws (PL) while HAWC J2031+415 is verified as a power law with an exponential cutoff (PLC) as predicted by VERITAS [4,6].  The simple power law equation is shown in Equation 1 and the power law with an exponential cutoff form is given in Equation 2.  

\begin{equation}
    \frac{dE}{dN} = N_o(\frac{E}{E_p})^{-\gamma}
\end{equation}

\begin{equation}
    \frac{dE}{dN} = N_o(\frac{E}{E_p})^{-\gamma}*\exp\frac{-E}{E_c}
\end{equation}

The free parameters are as follows: 3HWC J2020+403 has flux ($N_o$) and index ($\gamma$), HAWC J2030+409 has flux ($N_o$),, index ($\gamma$), and Gaussian width ($\sigma$), and HAWC J2031+415 has flux ($N_o$), index ($\gamma$), $\sigma$, and the cutoff ($E_c$) [4,12].  The pivot energies ($E_p$) are 1.1 TeV for 3HWC J2020+403, 4.2 TeV HAWC J2030+409, and 4.9 TeV for HAWC J2031+415 [4].  Once the fit is completed, the residual map displays no significant emission.  

To isolate HAWC J2031+415, the best fit parameters found for the Cocoon and Gamma Cygni are then fixed in the next iteration of the fitting process.  This is to subtract out their emission and leave only HAWC J2031+415, and this is shown in Figure 1b.  HAWC J2031+415 is then fitted again and compared to the 3 source fitting results to verify the validity of the subtraction method.  

\section{Morphology study}

Now that HAWC J2031+415 has been properly modelled, an investigation of its energy morphology is done.  It has been shown that the HAWC array does not have a large enough data set or adequate angular resolution to perform well constrained fits with individual bins or energy bands for bright sources as found in these systematic studies [5,13].  As such, a slicing method is considered.  Broadly speaking, this method takes a rectangular region of data and slices into a specified number of bins and counts the excess events.  Each slice effectively creates a smaller rectangle inside the larger rectangular region considered .  For this analysis, 50 excess event bins are considered to maximize the bin number while avoiding pixelation issues that can arise from slicing HEALPIX pixels too finely [5,13].  These bins will henceforth be referred to as count bins.  These count bins are then plotted in longitudinal profiles and a 1D Gaussian is fitted to the trend the count bins trace.  The Gaussian width is then treated as the width and is defined as PSF' for the 68$\%$ containment $\sigma$.  This $\sigma$ is approximately the width of the source, though comparison to simulations is required [5,13].

For HAWC J2031+415, the data set considered is the subtracted set as referenced in Section 3.1.  To maximize the data included in the energy bands, 4 bands will be used with the following reconstructed energy ranges: Band 1 (1 - 3.16 TeV), Band 2 (3.16 -10 TeV), Band 3 (10 - 56.2 TeV), and Band 4 (56.2 -316 TeV).  These correspond to ground parameter bins c d, e f, g h i, and j k l respectively [3].  As has been previously found, not all the bins have enough data due to the energy binning scheme and corresponding PSF' and so a combination of point source simulations and data comparisons must be done to reduce the total bin number to maximize data and minimize PSF' [3,13].  To do this, both the Crab Nebula and HAWC J2031+415 are simulated as simple PL and point sources at their respective known locations.  This is to also observe whether there is a significant difference with PSF' and declination.  The results are shown in Figure 2.

\begin{figure}[h!]
\centering
\begin{subfigure}{.45\textwidth}
  \centering
  \includegraphics[width=1\linewidth]{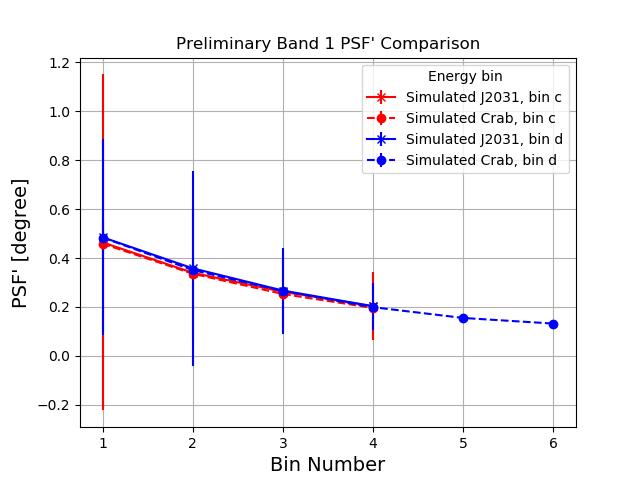}
  \caption{Band 1 (1 - 3.16 TeV).  This band contains all energy bins c and d.}
  \label{fig:sub1}
\end{subfigure}\hfill
\begin{subfigure}{.45\textwidth}
  \centering
  \includegraphics[width=1.0\linewidth]{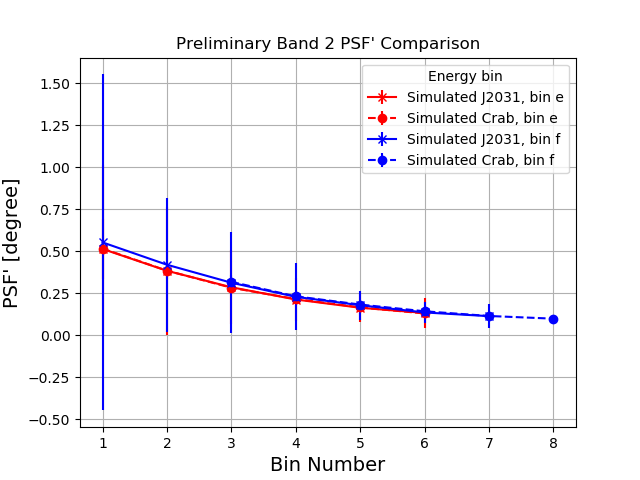}
  \caption{Band 2 (3.16 - 10 TeV).  This band contains energy bins e and f.}
  \label{fig:sub1}
\end{subfigure}\hfill
\begin{subfigure}{.45\textwidth}
  \centering
  \includegraphics[width=1.0\linewidth]{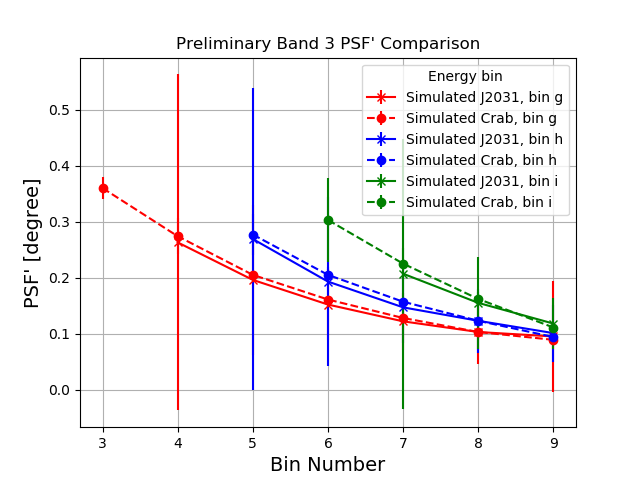}
  \caption{Band 3 (10 -56.2 TeV).  This band contains energy bins g, h, and i.}
  \label{fig:sub1}
\end{subfigure}\hfill
\begin{subfigure}{.45\textwidth}
  \centering
  \includegraphics[width=1.0\linewidth]{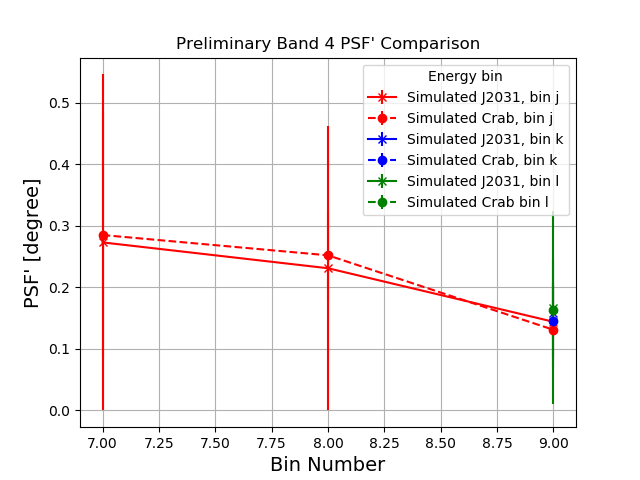}
  \caption{Band 4 (56.2 - 316 TeV).  This band contains energy bins j, k, and l.}
  \label{fig:sub1}
\end{subfigure}\hfill
\caption{Comparison of PSF' of HAWC J2031+415 and Crab simulations.  The x-axis represents each fhit bin that has a corresponding ground parameter energy bin (1c, 2c, etc).  The y-axis is the PSF' of each count bin.  As it can be see, the simulated PSF' of both sources very closely match and can be approximated as equivalent.}
\label{fig:test}
\end{figure}

It can be seen that there is no significant declination dependence with PSF' and as such the simulated Crab bins will be treated to determine the best count bins for each band.  The combined energy bin bands can be seen in Figure 3.  Here all energy bins corresponding to each fhit bin are combined.  The signal/noise (S/N) ratio is determined by dividing excess counts by the background.  

\begin{figure}[t]
\centering
\begin{subfigure}{.45\textwidth}
  \centering
  \includegraphics[width=1\linewidth]{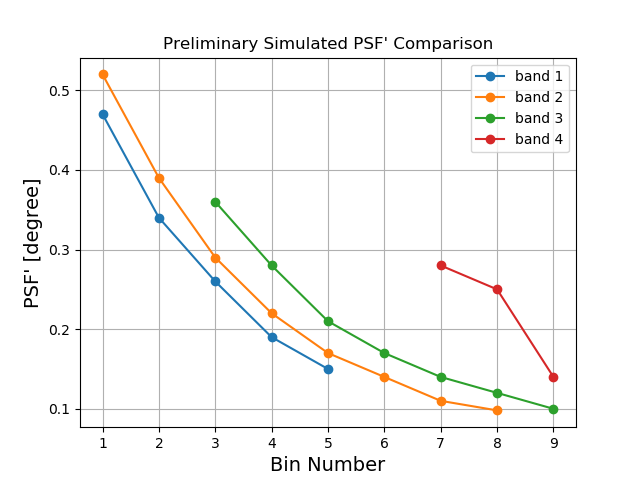}
  \caption{Comparing PSF' of each combined fhit bin of each energy band}
  \label{fig:sub1}
\end{subfigure}\hfill
\begin{subfigure}{.45\textwidth}
  \centering
  \includegraphics[width=1\textwidth]{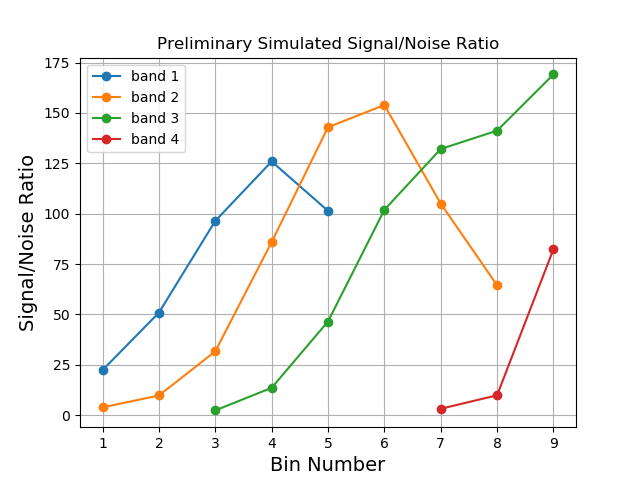}
  \caption{Comparing the S/N ratio of each combined fhit bin of each energy band.}
  \label{fig:sub2}
\end{subfigure}
\caption{Determining the best bins to keep for each energy band.  Each band is color-coded and corresponds to their respective energy intervals.}
\label{fig:test}
\end{figure}

The bins kept correspond to the highest S/N value and any that follow that bin, if applicable.  If bins with lower S/N values are kept, the PSF' should be roughly 75$\%$ of the highest S/N PSF', though exceptions can be made.  For example, Band 1 contains bins 3 and 4.  Bin 5 is not included as there is insufficient data at HAWC J2031+415's location.  With these bands a proper morphology study can be conducted.  The results are inconclusive, there is no significant trend towards decreasing energy morphology with increasing energy bands of HAWC J2031+415. 

\section{Discussion and conclusions}

This is the first in-depth high energy analysis of HAWC J2031+415.  The morphology of this source agrees with previous work within error bars and is detected in the energy spectrum to 100 TeV [4].  Considering the morphology study, issues were encountered with the amount of statistics contained in each band.  As the energy increased, there became too little data to adequately constrain the $\sigma$ of the fit.   However, this will be addressed soon as the outrigger array addressed earlier is being brought online.  This array will significantly increase HAWC's sensitivity of >10 TeV events.  This will greatly aid with this morphology study and may reveal additional shifts not currently visible.  Additionally, further work from the LHAASO observatory could reveal additional structure to this region.  With the current data set, HAWC J2031+415 is not visible above 100 TeV but the high energy detection by LHAASO in this region will be considered in future work [14]. 

\acknowledgments
We acknowledge the support from: the US National Science Foundation (NSF); the US Department of Energy Office of High-Energy Physics; the Laboratory Directed Research and Development (LDRD) program of Los Alamos National Laboratory; Consejo Nacional de Ciencia y Tecnolog\'ia (CONACyT), M\'exico, grants 271051, 232656, 260378, 179588, 254964, 258865, 243290, 132197, A1-S-46288, A1-S-22784, c\'atedras 873, 1563, 341, 323, Red HAWC, M\'exico; DGAPA-UNAM grants IG101320, IN111716-3, IN111419, IA102019, IN110621, IN110521; VIEP-BUAP; PIFI 2012, 2013, PROFOCIE 2014, 2015; the University of Wisconsin Alumni Research Foundation; the Institute of Geophysics, Planetary Physics, and Signatures at Los Alamos National Laboratory; Polish Science Centre grant, DEC-2017/27/B/ST9/02272; Coordinaci\'on de la Investigaci\'on Cient\'ifica de la Universidad Michoacana; Royal Society - Newton Advanced Fellowship 180385; Generalitat Valenciana, grant CIDEGENT/2018/034; Chulalongkorn University’s CUniverse (CUAASC) grant; Coordinaci\'on General Acad\'emica e Innovaci\'on (CGAI-UdeG), PRODEP-SEP UDG-CA-499; Institute of Cosmic Ray Research (ICRR), University of Tokyo, H.F. acknowledges support by NASA under award number 80GSFC21M0002. We also acknowledge the significant contributions over many years of Stefan Westerhoff, Gaurang Yodh and Arnulfo Zepeda Dominguez, all deceased members of the HAWC collaboration. Thanks to Scott Delay, Luciano D\'iaz and Eduardo Murrieta for technical support.

\clearpage
\section*{Full Authors List: \Coll\ Collaboration}


\scriptsize
\noindent
A.U. Abeysekara$^{48}$,
A. Albert$^{21}$,
R. Alfaro$^{14}$,
C. Alvarez$^{41}$,
J.D. Álvarez$^{40}$,
J.R. Angeles Camacho$^{14}$,
J.C. Arteaga-Velázquez$^{40}$,
K. P. Arunbabu$^{17}$,
D. Avila Rojas$^{14}$,
H.A. Ayala Solares$^{28}$
% ...

\noindent
% ...
$^{14}$Instituto de Física, Universidad Nacional Autónoma de México, Ciudad de México, México,
% ...
$^{17}$Instituto de Geofísica, Universidad Nacional Autónoma de México, Ciudad de México, México,
% ...
$^{21}$Physics Division, Los Alamos National Laboratory, Los Alamos, NM, USA,
% ...
$^{28}$Department of Physics, Pennsylvania State University, University Park, PA, USA,
% ...
$^{40}$Instituto de Física y Matemáticas, Universidad Michoacana de San Nicolás de Hidalgo, Morelia, Michoacán, México,
$^{41}$FCFM-MCTP, Universidad Autónoma de Chiapas, Tuxtla Gutiérrez, Chiapas, México,
% ...
$^{48}$Department of Physics and Astronomy, University of Utah, Salt Lake City, UT, USA,
% ...

%
%
%

\end{document}